\documentclass[conference]{IEEEtran}
\IEEEoverridecommandlockouts

\usepackage{cite}
\usepackage{amsmath,amssymb,amsfonts}
\usepackage{algorithmic}
\usepackage{graphicx}
\usepackage{textcomp}
\usepackage{xcolor}

\usepackage{subcaption}
\usepackage{bm}

\def\BibTeX{{\rm B\kern-.05em{\sc i\kern-.025em b}\kern-.08em
    T\kern-.1667em\lower.7ex\hbox{E}\kern-.125emX}}
\begin{document}

\title{Weighted Sum-Rate Maximization for RIS-UAV-assisted Space-Air-Ground Integrated Network with RSMA\\
}

\author{
	\IEEEauthorblockN{Jian He$^{*}$, Cong Zhou$^{*}$, Shuo Shi$^*$}
	\IEEEauthorblockA{$^*$ School of Electronics and Information Engineering, Harbin Institute of Technology, Harbin, China}
}

\maketitle

\begin{abstract}
In this paper, a rate-splitting multiple access (RSMA) based joint optimization framework for the space–air–ground integrated network (SAGIN) is proposed, where the satellite and base stations employ uniform planar array (UPA) antennas for signal transmission, and unmanned aerial vehicles (UAVs) relay the satellite signals. Earth stations (ESs) and user equipments (UEs) receive signals from satellite and base stations (BSs), respectively, resulting in mutual interference. We first model the channels and signals in this scenario and analyse the interference at BSs and UEs. Then, We formulate a joint optimization problem aimed at maximizing the weighted sum-rate, involving beamforming, RIS-UAV deployment and phase shifts, and rate splitting. However, this problem is highly non-convex. To tackle this challenge, we apply a block coordinate descent (BCD) approach to decompose the problem and employ the weighted minimum mean square error (WMMSE) method to transform the non-convex objective function. For the rate-splitting sub-problem, a greedy algorithm is proposed and a successive convex approximation (SCA) algorithm is used for beamforming. Besides, the alternating direction method of multipliers (ADMM) algorithm is employed for the RIS phase-shift problem with unit-modulus constraints, and an exhaustive search method is adopted for the complex UAV positioning and orientation. Simulation results validate that the proposed algorithm achieves superior performance in terms of user weighted sum-rate.
\end{abstract}

\section{Introduction}
With the rapid growth of global communication services, the demand for ubiquitous, high-data-rate, and low-latency communications has been continuously increasing \cite{intro1-6G}. 
However, traditional terrestrial communication networks are constrained by geographical environments, infrastructure limitations, and natural disasters, making it difficult to achieve seamless coverage in complex or remote areas \cite{intro2-satellite}\cite{zhou2024energy}.
To address these challenges, the space–air–ground integrated network (SAGIN) \cite{intro3-SAGIN} has been recognized as one of the key architectures for 6G. By integrating satellite, aerial, and terrestrial networks, SAGIN enables global, three-dimensional, and intelligent communication connectivity \cite{intro4-SAGIN}. In this architecture, terrestrial networks handle high-density and high-data-rate demands, space-based networks provide global coverage, and aerial platforms offer dynamic coverage compensation. 
However, the integration of multi-layer heterogeneous systems in SAGIN also introduces significant challenges in resource coordination and interference management. The dynamic mobility of aerial platforms and the complexity of communication channels further increase the difficulty of system design and optimization \cite{intro5-SAGINEC}.

Among various enabling technologies, unmanned aerial vehicles (UAVs) have emerged as key relay nodes in SAGIN due to their flexible deployment and controllable mobility, which enhance coverage and link quality \cite{10238400}. In recent years, the concept of integrating reconfigurable intelligent surfaces (RIS) with UAVs, forming a RIS-UAV architecture, has attracted growing attention \cite{intro10-RIS-UAV}. By adjusting the phase shifts of its reflecting elements, the RIS can intelligently reconfigure the propagation environment \cite{intro11-RIS}, while the UAV’s three-dimensional mobility allows dynamic adjustment of the reflection position, thereby establishing virtual line-of-sight (LoS) links in complex terrains.
Meanwhile, rate-splitting multiple access (RSMA), as a novel multiple access scheme, can effectively manage multi-user interference and improve spectral efficiency through message splitting and partial interference decoding \cite{intro13-RSMA}. Incorporating RSMA into SAGIN can significantly enhance system performance. However, it also introduces strong coupling among beamforming design, UAV deployment, and RIS phase-shift optimization, leading to a highly non-convex problem. This paper investigates a joint optimization framework for satellite and terrestrial beamforming, RIS-UAV three-dimensional deployment, and RSMA rate allocation to enhance the overall performance of the communication system.
\section{SYSTEM MODEL}
\begin{figure}[!t]
	\centering
	\includegraphics[width=2.5in]{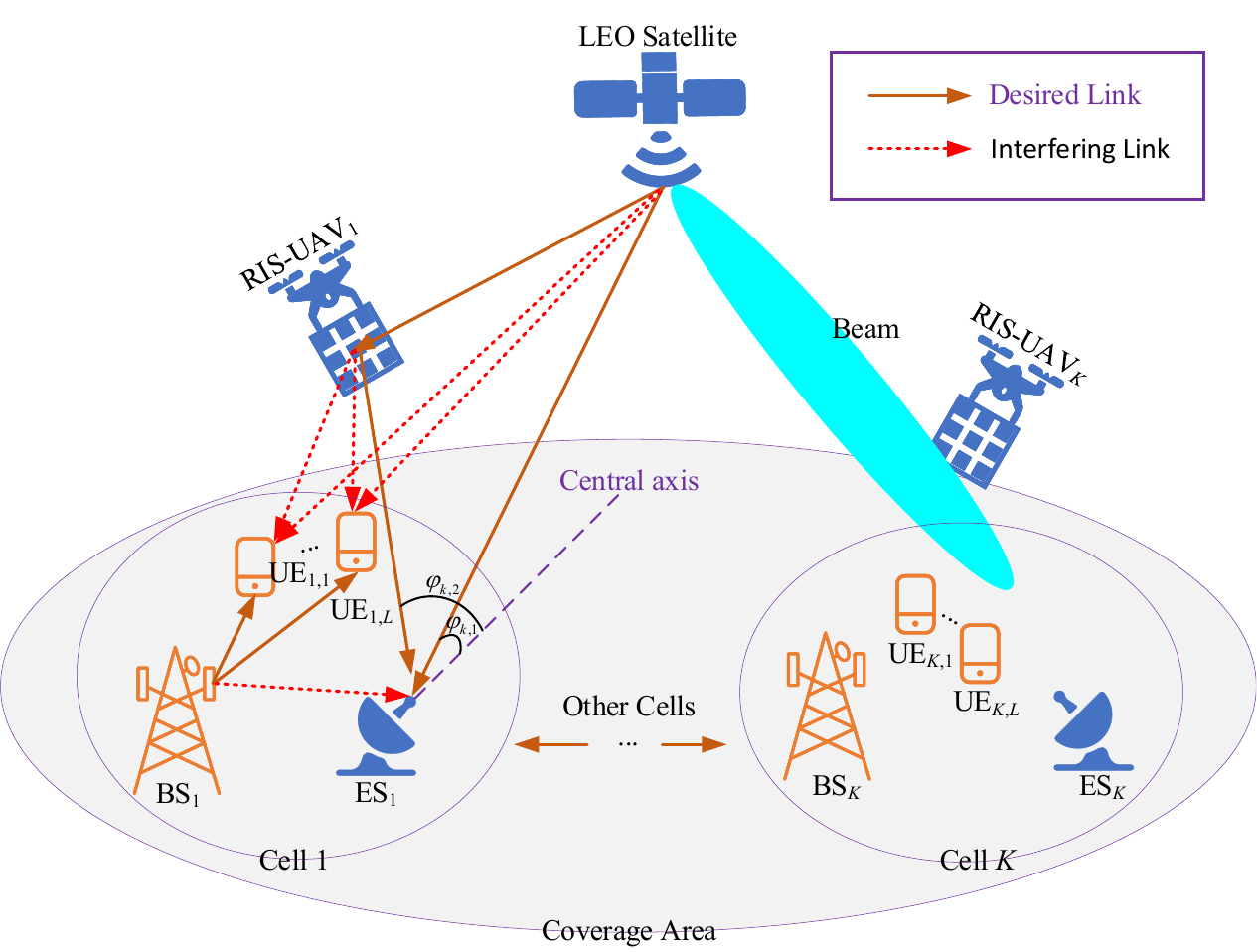}
	\caption{System model of considered SAGIN.}
	\label{model}
	\vspace*{-16pt}
\end{figure}

The system model is illustrated in the Fig.\ref{model}. Within the coverage area of a low earth orbit (LEO) satellite, there are numerous terrestrial cells, each comprising a BS, an earth station (ES), an RIS-UAV and multiple UEs. The LEO satellite serves the ESs, each of which is assisted by a dedicated RIS-UAV to enhance the signal quality. Meanwhile, the BS equipped with a UPA, using the same frequency $f$ as the LEO satellite, provides downlink services to the UEs. Moreover, the LEO satellite is also equipped with the UPA.
\vspace*{-4pt}
\subsection{Channel Model}
\underline{\textbf{Satellite downlink model: }}In the downlink, the LEO satellite, located at an altitude of $H_{\text{S}}$ and consisting of $N_{\text{S}}=N_{\text{S}x} \times N_{\text{S}y}$ antennas, serves $K$ ESs within the same time slot. Hence, with $k \in \mathcal{K} \triangleq \{1,2,...,K\}$, the channel between the LEO satellite and $k$-th ES can be expressed as
\begin{equation} \label{channalS2E}
	\mathbf{h}_k = \sqrt{\frac{N_{\text{S}} G(\varphi_{k,1})}{\beta_{k} \xi_{k}}} \cdot \mathbf{a}_{\text{S}}(\theta^{\text{ES}}_{\text{S},k},\phi^{\text{ES}}_{\text{S},k}), \forall k \in \mathcal{K},
\end{equation}
where $\beta_{k}$ and $\xi_{k}$ denote the free-space path loss and rain attenuation, respectively. The $\beta_{k}$ can be obtained using the expression $\beta_{k} = (4\pi d^{\text{ES}}_{k}/\lambda)^2$ with $d^{\text{ES}}_{k}$ and $\lambda$ denoting the distance from the satellite to the $k$-th ES and the wavelength, respectively. The dB form of $\xi_{k}$ follows a log-normal distribution $\ln(\xi^{\text{dB}}_k) \sim \mathcal{N}(\mu, \sigma^2)$ \cite{model-rain_attenuation}. In addition, $G(\varphi_{k,1})$ in (\ref{channalS2E}) denotes the receive antenna gain at the ES as \cite{model-G_r}
\begin{equation}
	\hspace*{-0.45em}
	\label{eq_receive_gain}
	G^{\text{dBi}}(\varphi) = \begin{cases}
		G^{\text{dBi}}_{\text{max}}-0.0025(\frac{D \cdot \varphi}{\lambda})^2, & \varphi \in [0,\varphi_{\text{m}})\\
		32-25\log(\varphi_{\text{r}}), & \varphi \in [\varphi_{\text{m}},\varphi_{\text{r}}]\\
		\max(32-25\log\varphi,-10), & \varphi \in (\varphi_{\text{r}},\pi]
	\end{cases},
\end{equation}
\noindent where $G_{\text{max}}=\eta(\pi D/\lambda)^2$ denotes the maximum antenna gain. In addition, $\varphi_{\text{m}} = 20(\lambda/D)\sqrt{G^{\text{dBi}}_{\text{max}}-32-25\log(\varphi_{\text{r}})}$ and $\varphi_{\text{r}} = 15.85(D/\lambda)^{-0.6}$.

Furthermore, $\mathbf{a}_{\text{S}}(\theta^{\text{ES}}_{\text{S},k}, \phi^{\text{ES}}_{\text{S},k}) \in \mathbb{C}^{N_{\text{S}} \times 1}$ in (\ref{channalS2E}), with $\theta^{\text{ES}}_{\text{S},k} \in [0,\pi /2]$ and $\phi^{\text{ES}}_{\text{S},k} \in [-\pi,\pi]$ denoting the elevation and azimuth angles of the ES, represents the steering vector
\begin{equation}\label{steering_vector}
	\mathbf{a}_{\text{S}}(\theta^{\text{ES}}_{\text{S},k},\phi^{\text{ES}}_{\text{S},k}) =\frac{1}{\sqrt{N_{\text{S}}}} \mathbf{a}_{\text{S},x}(\theta^{\text{ES}}_{\text{S},k},\phi^{\text{ES}}_{\text{S},k}) \otimes \mathbf{a}_{\text{S},y}(\theta^{\text{ES}}_{\text{S},k},\phi^{\text{ES}}_{\text{S},k}),
\end{equation}
where $\mathbf{a}_{\text{S},x}(\theta^{\text{ES}}_{\text{S},k},\phi^{\text{ES}}_{\text{S},k}) \in \mathbb{C}^{N_{\text{Sx}}\times 1} $ and $\mathbf{a}_{\text{S},y}(\theta^{\text{ES}}_{\text{S},k},\phi^{\text{ES}}_{\text{S},k}) \in \mathbb{C}^{N_{\text{Sy}}\times 1} $ are channel response vectors in two directions, respectively. Moreover, with $\mathcal{N}_x \triangleq \{1,2,...,N_{\text{S}x}\}$ and $\mathcal{N}_y \triangleq \{1,2,...,N_{\text{S}y}\}$, the elements of them can be expressed as:
\begin{equation}\label{steering_vector_x}
	\small
	[\mathbf{a}_{\text{S},x}(\theta^{\text{ES}}_{\text{S},k},\phi^{\text{ES}}_{\text{S},k})]_{n_{x}} = e^{\jmath \frac{2\pi}{\lambda} d_0 (n_{x}-1) \sin\theta^{\text{ES}}_{\text{S},k} \cos\phi^{\text{ES}}_{\text{S},k}}, \forall n_{x} \in \mathcal{N}_x ,
\end{equation}
\begin{equation}\label{steering_vector_y}
	\small
	[\mathbf{a}_{\text{S},y}(\theta^{\text{ES}}_{\text{S},k},\phi^{\text{ES}}_{\text{S},k})]_{n_{y}} = e^{\jmath \frac{2\pi}{\lambda} d_0 (n_{x}-1) \sin\theta^{\text{ES}}_{\text{S},k} \sin\phi^{\text{ES}}_{\text{S},k}}, \forall n_{y} \in \mathcal{N}_y,
\end{equation}
where $d_0$ represents the inter-antenna spacing. Furthermore, the channel from the satellite to the $k$-th RIS-UAV with height $H_{\text{R}}$ can be represented by $\mathbf{G}_k \in \mathbb{C}^{N_{\text{R}}\times N_{\text{S}}} $ with $N_{\text{R}} = N_{\text{R}x} \times N_{\text{R}y}$ denoting the number of elements of the RIS. In addition, denoting the channel from the $k$-th RIS-UAV to the ES and the RIS phase shift matrix by $\mathbf{g}_{k} \in \mathbb{C}^{N_{\text{R}} \times 1}$ and $\mathbf{\Psi}_{k} \in \mathbb{C}^{N_{\text{R}} \times N_{\text{R}}}$, the equivalent channel $\tilde{\mathbf{h}}_k \in \mathbb{C}^{N_{\text{S}} \times 1}$ can be expressed as
\begin{equation}
	\tilde{\mathbf{h}}^{H}_k = \mathbf{h}^{H}_k + \mathbf{g}^{H}_{k} \mathbf{\Psi}_{k} \mathbf{G}_k, \forall k \in \mathcal{K},
\end{equation}
where $\mathbf{\Psi}_{k} = \mathrm{diag}(e^{\jmath \psi_{k,1}},e^{\jmath \psi_{k,2}},...e^{\jmath \psi_{k,N_{\text{R}}}})$. Besides, we assume that the $k$-th UAV is aligned with the central axis of the parabolic antenna at the $k$-th ES, i.e., $\varphi_{k,2}=0$.

\underline{\textbf{BS downlink model: }}The BS equipped with $N_{\text{B}}=N_{\text{B}x} \times N_{\text{B}y}$ antennas UPA in the $k$-th cell provides services to $L$ UEs with single antenna in the same time slot. Particularly, we denote the height of UPA by $H_{\text{B}}$ and assume that all UEs serviced by the $k$-th BS are at the same height as the $k$-th ES. Moreover, with $\mathcal{L} \triangleq \{1,2,...,L\}$, the elevation and azimuth angles of the $l$-th UE relative to the UPA of BS are denoted by $\theta^{\text{U}}_{k,l}$ and $\phi^{\text{U}}_{k,l}$, respectively. Hence, the channel from the BS to the $l$-th UE $\mathbf{v}_{k,l} \in \mathbb{C}^{N_{\text{B}} \times 1}$ can be expressed as
\vspace*{-6pt}
\begin{equation}
	\mathbf{v}_{k,l} = \frac{\lambda \sqrt{N_{\text{B}}}}{4 \pi d^{U}_{k,l}} \cdot \mathbf{b}(\theta^{\text{U}}_{k,l},\phi^{\text{U}}_{k,l}), \forall k \in \mathcal{K}, l \in \mathcal{L},
\end{equation}
where $d^{\text{U}}_{k,l}$ represents the distance from the BS to $l$-th UE.

There are two ways to ignore the energy leakage from the BS to RIS-UAV: we can deploy the UAV at a sufficiently high altitude, or orient the RIS reflection surface toward the ES while placing the BS on the backside of the RIS.

\underline{\textbf{Interfering link model: }}We use $\mathbf{f}_{k,l}$, $\mathbf{q}_{k,l}$ and $\mathbf{u}_{k}$ to represent the channel gain from the satellite to UE, from RIS-UAV to UE and from the BS to the ES, and it is obvious that they are similar to $\mathbf{h}_{k}$, $\mathbf{g}_{k}$ and $\mathbf{v}_{k,l}$, respectively.

Assuming that the array plane at the satellite lies in the x-y plane which is parallel to the ground plane, the coordinate of the $k$-th ES is given by $\mathbf{Q}^{\text{E},\mathcal{S}_{0}}_{k}=(x^{\text{E}}_{k},y^{\text{E}}_{k},z^{\text{E}}_{k})^{T}$ in this 3D Cartesian coordinate system defined by $\mathcal{S}_{0}$.

Analogously, the coordinate of $l$-th UE of $k$-th cell in $\mathcal{S}^{\text{B}}_{k}$ is given by $\mathbf{Q}^{\text{U},\mathcal{S}^{\text{B}}_{k}}_{k,l}=(x^{\text{U}}_{k,l},y^{\text{U}}_{k,l},z^{\text{U}}_{k,l})^{T}$.
Similarly, the coordinate system $\mathcal{S}^{\text{R}}_{k}$ is defined such that the $k$-th RIS lies in the x-y plane, with its facing direction aligned with the z-axis. Furthermore, giving the rotation matrix and translation vector from $\mathcal{S}_{0}$ to $\mathcal{S}^{\text{R}}_{k}$ by $\mathbf{R}^{\text{R}}_k \in \mathbf{R}^{\text{R}} \triangleq \{\mathbf{R}^{\text{R}}_k\}^{K}_{k=1}$ and $\mathbf{t}^{\text{R}}_k \in \mathbf{t}^{\text{R}} \triangleq {\{\mathbf{t}^{\text{R}}_k\}^{K}_{k=1}}$, respectively, the elevation and azimuth angle can be obtained to calculate $\mathbf{g}_{k}$, $\mathbf{G}_{k}$, $\mathbf{f}_{k,l}$, $\mathbf{q}_{k,l}$ and $\mathbf{u}_{k}$.
\subsection{Signal Model}
Both the satellite and the BSs employ RSMA for signal transmission. Denoting common and private steam by $s^{\text{S}}_{\text{c}}$ and $s^{\text{S}}_k$, the signal transmitted by the satellite can be expressed as
\vspace*{-6pt}
\begin{equation}
	\mathbf{x}^{\text{S}} = \mathbf{w}^{\text{S}}_{\text{c}}s^{\text{S}}_{\text{c}}+\sum_{k=1}^{K} \mathbf{w}^{\text{S}}_k s^{\text{S}}_k, \forall k \in \mathcal{K} \triangleq {1,2,...,K},
\end{equation}
where the $\mathbf{w}^{\text{S}}_{\text{c}}$ and $\mathbf{w}^{\text{S}}_k \in \mathbb{C}^{N_{\text{S}} \times 1}$ represent the beamforming vector. Similarly, the signal of the $k$-th BS can be given by
\vspace*{-6pt}
\begin{equation}
	\mathbf{x}^{\text{B}}_k = \mathbf{w}^{\text{B}}_{k,\text{c}}s^{\text{B}}_{k,\text{c}}+\sum_{l=1}^{L} \mathbf{w}^{\text{B}}_{k,l} s^{\text{B}}_{k,l}, \forall k \in \mathcal{K},
\end{equation}
where the $\mathbf{w}^{\text{B}}_{k,\text{c}}$ and $\mathbf{w}^{\text{B}}_{k,l} \in \mathbb{C}^{N_{\text{B}} \times 1}$ denote the beamforming vector at the $k$-th BS. With $\tilde{\mathbf{f}}^{H}_{k,l}$ denoting the channel from satellite to $l$-th UE as (\ref{channel_satellite_UE}), the received signal at the ES and UE can be expressed as (\ref{ES_receive_signal}) and (\ref{UE_receive_signal}), where $n^{\text{E}}_{k}$ and $n^{\text{U}}_{k,l}$ denotes the Additive White Gaussian Noise (AWGN).
\begin{equation}
	\label{channel_satellite_UE}
	\tilde{\mathbf{f}}^{H}_{k,l} = \mathbf{f}^{H}_{k,l} + \mathbf{q}^{H}_{k,l} \mathbf{\Psi}_k \mathbf{G}_k.
\end{equation}
\begin{figure*}[t]
	\normalsize
	\begin{equation}
		\label{ES_receive_signal}
		\begin{aligned}
			y^{\text{E}}_k = \tilde{\mathbf{h}}^{H}_k \mathbf{x}^{\text{S}} + \mathbf{u}^{H}_{k} \mathbf{x}^{\text{B}}_{k} + n^{\text{E}}_{k} = \underbrace{\tilde{\mathbf{h}}^{H}_{k} \mathbf{w}^{\text{S}}_{\text{c}} s^{\text{S}}_{\text{c}}}_{\text{common signal}} + \underbrace{\tilde{\mathbf{h}}^{H}_{k} \mathbf{w}^{\text{S}}_{k}s^{\text{S}}_{k}}_{\text{private signal}} + \underbrace{\tilde{\mathbf{h}}^{H}_{k} \sum_{j \neq k}^{K} \mathbf{w}^{\text{S}}_j s^{\text{S}}_j}_{\text{multi-ES interference}}  + \underbrace{\mathbf{u}^{H}_{k} \mathbf{w}^{\text{B}}_{k,\text{c}} s^{\text{B}}_{k,\text{c}} + \mathbf{u}^{H}_{k} \sum_{j=1}^{L} \mathbf{w}^{\text{B}}_{k,j} s^{\text{B}}_{k,j}}_{\text{interference from $k$-th BS}}  + n^{\text{E}}_{k}.
		\end{aligned}
	\end{equation}
	\vspace*{-6pt}
	\begin{equation}
		\label{UE_receive_signal}
		\begin{aligned}
			y^{\text{U}}_{k,l} = \mathbf{v}^{H}_{k,l} \mathbf{x}^{\text{B}}_{k} + \tilde{\mathbf{f}}^{H}_{k,l} \mathbf{x}^{\text{S}} + n^{\text{U}}_{k,l} = \underbrace{\mathbf{v}^{H}_{k,l} \mathbf{w}^{\text{B}}_{k,\text{c}} s^{\text{B}}_{k,\text{c}}}_{\text{common signal}} + \underbrace{\mathbf{v}^{H}_{k,l} \mathbf{w}^{\text{B}}_{k,l} s^{\text{B}}_{k,l}}_{\text{private signal}} + \underbrace{\mathbf{v}^{H}_{k,l} \sum_{j \neq l}^{L} \mathbf{w}^{\text{B}}_{k,j} s^{\text{B}}_{k,j}}_{\text{multi-UE interference}} + \underbrace{\tilde{\mathbf{f}}^{H}_{k,l} \mathbf{w}^{\text{S}}_{\text{c}} s^{\text{S}}_{\text{c}} + \tilde{\mathbf{f}}^{H}_{k,l} \sum_{j=1}^{K} \mathbf{w}^{\text{S}}_{j} s^{\text{S}}_{j}}_{\text{interference from satellite}} + n^{\text{U}}_{k,l}.
		\end{aligned}
	\end{equation}
	\hrulefill
	\vspace*{-16pt}
\end{figure*}
With $\mathbb{E}\{|s^{\text{S}}_{k}|^2\}=1$ and $\mathbb{E}\{|s^{\text{B}}_{k,l}|^2\}=1$, the associated signal-to-interference-plus-noise ratio (SINR) can be expressed as
\begin{equation}
	\small
	\gamma^{\text{E,c}}_{k} = \frac{|\tilde{\mathbf{h}}^{H}_{k} \mathbf{w}^{\text{S}}_{\text{c}}|^2}{\sum\limits_{j=1}^{K}|\tilde{\mathbf{h}}^{H}_{k} \mathbf{w}^{\text{S}}_{j}|^{2} + |\mathbf{u}^{H}_{k} \mathbf{w}^{\text{B}}_{k,c}|^{2} + \sum\limits_{j=1}^{L}|\mathbf{u}^{H}_{k} \mathbf{w}^{\text{B}}_{k,j}|^{2} + P^{\text{E}}_{k}},
\end{equation}
\begin{equation}
	\small
	\gamma^{\text{E,p}}_{k} = \frac{|\tilde{\mathbf{h}}^{H}_{k} \mathbf{w}^{\text{S}}_{k}|^2}{\sum\limits_{j \neq k}^{K}|\tilde{\mathbf{h}}^{H}_{k} \mathbf{w}^{\text{S}}_{j}|^{2} + |\mathbf{u}^{H}_{k} \mathbf{w}^{\text{B}}_{k,c}|^{2} + \sum\limits_{j=1}^{L}|\mathbf{u}^{H}_{k} \mathbf{w}^{\text{B}}_{k,j}|^{2} + P^{\text{E}}_{k}},
\end{equation}
\begin{equation}
	\small
	\gamma^{\text{U,c}}_{k,l} = \frac{|\mathbf{v}^{H}_{k,l} \mathbf{w}^{\text{B}}_{k,\text{c}}|^2}{\sum\limits_{j=1}^{L}|\mathbf{v}^{H}_{k,l} \mathbf{w}^{\text{B}}_{k,j}|^{2} + |\tilde{\mathbf{f}}^{H}_{k,l} \mathbf{w}^{\text{S}}_{c}|^{2} + \sum\limits_{j=1}^{K}|\tilde{\mathbf{f}}^{H}_{k,l} \mathbf{w}^{\text{S}}_{j}|^{2} + P^{\text{U}}_{k,l}},
\end{equation}
\begin{equation}
	\small
	\gamma^{\text{U,p}}_{k,l} = \frac{|\mathbf{v}^{H}_{k,l} \mathbf{w}^{\text{B}}_{k,l}|^2}{\sum\limits_{j \neq l}^{L}|\mathbf{v}^{H}_{k,l} \mathbf{w}^{\text{B}}_{k,j}|^{2} + |\tilde{\mathbf{f}}^{H}_{k,l} \mathbf{w}^{\text{S}}_{c}|^{2} + \sum\limits_{j=1}^{K}|\tilde{\mathbf{f}}^{H}_{k,l} \mathbf{w}^{\text{S}}_{j}|^{2} + P^{\text{U}}_{k,l}},
\end{equation}
where $P^{\text{E}}_{k}$ and $P^{\text{U}}_{k,l}$ denote the noise power. Moreover, the rates of common streams are represented by $r^{\text{E}}_{k}$ and $r^{\text{U}}_{k,l}$, and the total rates of the ES and UE can be expressed as
\begin{equation}
	R^{\text{E}}_{k} = r^{\text{E}}_{k} + \log_{2}(1+\gamma^{\text{E,p}}_{k}),
\end{equation}
\begin{equation}
	R^{\text{U}}_{k,l} = r^{\text{U}}_{k,l} + \log_{2}(1+\gamma^{\text{U,p}}_{k,l}).
\end{equation}

\subsection{Problem Formulation}
We assign the weight as $\sum_{k=1}^{K}\sum_{l=0}^{L}\alpha_{k,l}=1$ and the problem can be formulated as
\begin{align}
	\hspace*{-0.45em}
	(\text{P1})\max_{\mathbf{\Xi}} \quad & \tilde{f}(\mathbf{\Xi})= \sum\limits_{k=1}^{K} \alpha_{k,0} R^{\text{E}}_{k} + \sum\limits_{k=1}^{K} \sum\limits_{l=1}^{L} \alpha_{k,l} R^{\text{U}}_{k,l} \tag{P1.a} \label{P1.a} \\
	\text{s.t.} \quad
	& r^{\text{E}}_{k}, r^{\text{U}}_{k,l} \geq 0, \forall k \in \mathcal{K}, \forall l \in \mathcal{L}, \tag{P1.b} \label{P1.b} \\
	& \sum\limits_{k=1}^{K} r^{\text{E}}_{k} \leq \min_{k} \log_{2}(1+\gamma^{\text{E,c}}_{k}), \tag{P1.c} \label{P1.c} \\
	& \sum\limits_{l=1}^{L} r^{\text{U}}_{k,l} \leq \min_{l} \log_{2}(1+\gamma^{\text{U,c}}_{k,l}), \forall k \in \mathcal{K}, \tag{P1.d} \label{P1.d} \\
	& \| \mathbf{w}^{\text{S}}_{\text{c}} \|^2 + \sum\limits_{k=1}^{K} \| \mathbf{w}^{\text{S}}_{k} \|^2 \leq P^{\text{S}}_{\text{max}}, \tag{P1.e} \label{P1.e} \\
	& \| \mathbf{w}^{\text{B}}_{k,\text{c}} \|^2 + \sum\limits_{l=1}^{L} \| \mathbf{w}^{\text{B}}_{k,l} \|^2 \leq P^{\text{B}}_{\text{max}}, \forall k \in \mathcal{K}, \tag{P1.f} \label{P1.f} \\
	& 0 \leq \psi_{k,j} < 2 \pi, \forall k \in \mathcal{K}, j=1,2,...,N_{\text{R}}, \tag{P1.g} \label{P1.g} \\
	& (\mathbf{R}^{\text{R}}_{k})^{T} \mathbf{R}^{\text{R}}_{k} = \mathbf{I}, \det(\mathbf{R}^{\text{R}}_{k})=1, \forall k \in \mathcal{K} \tag{P1.h} \label{P1.h} \\
	&  \mathbf{Q}^{\text{R},\mathcal{S}_{0}}_{k} \in \mathcal{X}_{k}, \forall k \in \mathcal{K}, \tag{P1.i} \label{P1.i} \\
	& [\mathbf{Q}^{\text{S},\mathcal{S}^{\text{R}}_{k}}]_{3}, [\mathbf{Q}^{\text{E},\mathcal{S}^{\text{R}}_{k}}_{k}]_{3} > 0, \forall k \in \mathcal{K}, \tag{P1.j} \label{P1.j} \\
	& R^{\text{E}}_{k} \geq R^{\text{E}}_{\text{min}}, R^{\text{U}}_{k,l} \geq R^{\text{U}}_{\text{min}}, \forall k \in \mathcal{K}, l \in \mathcal{L}. \tag{P1.k} \label{P1.k}
\end{align}
where $\mathbf{\Xi} \triangleq \{ \mathbf{w}^{\text{S}}, \mathbf{w}^{\text{B}}, \mathbf{\Psi}, \mathbf{r}^{\text{E}}, \mathbf{r}^{\text{U}}, \mathbf{R}^{\text{R}}, \mathbf{t}^{\text{R}} \}$ with $\mathbf{w}^{\text{S}} \triangleq (\mathbf{w}^{\text{S}}_{\text{c}},\mathbf{w}^{\text{S}}_{1},...\mathbf{w}^{\text{S}}_{K})$, $\mathbf{w}^{\text{B}} \triangleq (\mathbf{w}^{\text{B}}_{1,\text{c}},\mathbf{w}^{\text{B}}_{1,1},...\mathbf{w}^{\text{B}}_{K,L})$, $\mathbf{\Psi} \triangleq (\mathbf{\Psi}_1,\mathbf{\Psi}_2,...,\mathbf{\Psi}_K)$, $\mathbf{r}^{\text{E}} \triangleq (\mathbf{r}^{\text{E}}_{1},\mathbf{r}^{\text{E}}_{2},...,\mathbf{r}^{\text{E}}_{K})$ and $\mathbf{r}^{\text{U}} \triangleq (\mathbf{r}^{\text{U}}_{1,1} \mathbf{r}^{\text{U}}_{1,2},...,\mathbf{r}^{\text{U}}_{K,L})$. Constrains (\ref{P1.b})(\ref{P1.c}) and (\ref{P1.d}) ensure that the rates of common streams are finite and non-negative. Additionally, (\ref{P1.e}) and (\ref{P1.f}) are the maximum transmit power constraints at the satellite and BSs, respectively. (\ref{P1.g}) corresponds to the phase constraint of the RIS. Moreover, constrain (\ref{P1.h}) ensures that $\mathbf{R}^{\text{R}}_{k}$ is a valid rotation matrix which preserves lengths and angles. (\ref{P1.i}) is the position constrain. Besides, (\ref{P1.j}) indicates that both the satellite and $k$-th ES are in the forward half-space. (\ref{P1.k}) is the QoS constrain of the ESs and UEs.
\section{PROPOSED ALGORITHM}
To resolve this complex problem, we employ the WMMSE approach to transform it into an equivalent form, and at the same time employ block coordinate descent (BCD) algorithm to split (P1) into several more manageable sub-problems, which can be solved by converting its objective function and constraints into convex forms.

\subsection{Optimizing $\{\mathbf{r}^{\textnormal{E}},\mathbf{r}^{\textnormal{U}}\}$ Given $\{\mathbf{w}^{\textnormal{S}}, \mathbf{w}^{\textnormal{B}}, \mathbf{\Psi},\mathbf{R}^{\textnormal{R}}, \mathbf{t}^{\textnormal{R}}\}$}
The sub-problem (P1.1) can be formulated as follows with $\{\mathbf{w}^{\text{S}}, \mathbf{w}^{\text{B}}, \mathbf{\Psi},\mathbf{R}^{\text{R}}, \mathbf{t}^{\text{R}}\}$ within the feasible region.
\begin{align}
	(\text{P1.1}) \max_{\mathbf{r}^{\text{E}},\mathbf{r}^{\text{U}}} &\sum\limits_{k=1}^{K} \alpha_{k,0} r^{\text{E}}_{k} + \sum\limits_{k=1}^{K} \sum\limits_{l=1}^{L} \alpha_{k,l} r^{\text{U}}_{k,l} \tag{P1.1.a} \label{P1.1.a} \\
	\text{s.t.} & (\text{\ref{P1.b}}),(\text{\ref{P1.c}}),(\text{\ref{P1.d}}),(\text{\ref{P1.k}}). \tag{P1.1.b} \label{P1.1.b}
\end{align}

To facilitate analysis, the constraints (\ref{P1.b}) and (\ref{P1.k}) are organized as
\begin{equation}
	r^{\text{E}}_{k} \geq \max\{0,R^{\text{E}}_{\text{min}} - \log_{2}(1+\gamma^{\text{E,p}}_{k})\}, \label{P1.1.c}
\end{equation}
\begin{equation}
	r^{\text{U}}_{k,l} \geq \max\{0,R^{\text{U}}_{\text{min}} - \log_{2}(1+\gamma^{\text{U,p}}_{k,l})\}.\label{P1.1.d}
\end{equation}

Hence, the problem (P1.1) with constrains (\ref{P1.c}), (\ref{P1.d}), (\ref{P1.1.c}) and (\ref{P1.1.d}) can be solved by a greedy algorithm as follows. First, the rate is allocated to each ES or UE to satisfy all the lower-bound constraints. Then, under the condition that the upper-bound constraints are met, the remaining rate is fully allocated to the ES or UE with the highest weight. 
\subsection{Equivalent Form of the Problem (P1)}
When optimizing the remaining variables, we define $\mathbf{\Upsilon} \triangleq \{\mathbf{w}^{\text{S}}, \mathbf{w}^{\text{B}}, \mathbf{\Psi},\mathbf{R}^{\text{R}}, \mathbf{t}^{\text{R}}\}$ and employ the WMMSE algorithm to transform problem (P1) into the following form \cite{WMMSE}.
\begin{align}
	(\text{P2})\min_{\mathbf{\Upsilon}, \bm{\mu}, \bm{\omega}} &\sum_{k=1}^{K} \sum_{j=0}^{L} \alpha_{k,j}(\omega_{k,j} e_{k,j} - \log(\omega_{k,j})) \tag{P2.a} \label{P2.a} \\
	\text{s.t.} &(\text{\ref{P1.e}})-(\text{\ref{P1.k}}), \tag{P2.b} \label{P2.b}
\end{align}
where $e_{k,j} = \mathbb{E}[|\mu_{k,j} y_{k,j} - s_{k,j}|^{2}]$, and can be expanded as
\begin{align}
	e_{k,0} = &|\mu_{k,0}|^{2} (\sum_{j=1}^{K}|\tilde{\mathbf{h}}^{H}_{k} \mathbf{w}^{\text{S}}_{j}|^{2} + |\mathbf{u}^{H}_{k} \mathbf{w}^{\text{B}}_{k,\text{c}}|^2 + \sum_{j=1}^{L}|\mathbf{u}^{H}_{k} \mathbf{w}^{\text{B}}_{k,j}|^{2} \notag \\
	&+ P^{\text{E}}_{k}) - 2 \text{Re}\{\mu_{k,0} \tilde{\mathbf{h}}^{H}_{k} \mathbf{w}^{\text{S}}_{k}\} + 1,
\end{align}
\begin{align}
	e_{k,l} = &|\mu_{k,l}|^{2} (\sum_{j=1}^{L}|\mathbf{v}^{H}_{k,l} \mathbf{w}^{\text{B}}_{k,j}|^{2} + |\tilde{\mathbf{f}}^{H}_{k,l} \mathbf{w}^{\text{S}}_{\text{c}}|^2 + \sum_{j=1}^{K}|\tilde{\mathbf{f}}^{H}_{k,l} \mathbf{w}^{\text{S}}_{j}|^{2} \notag \\
	&+P^{\text{U}}_{k,l}) - 2 \text{Re}\{\mu_{k,l} \mathbf{v}^{H}_{k,l} \mathbf{w}^{\text{B}}_{k,l}\} + 1, \forall l \in \mathcal{L}.
\end{align}

In each optimization iteration, we update $\bm{\mu}$ and $\bm{\omega}$:
\begin{equation}
	\label{update_mu1}
	\small
	\mu^{*}_{k,0} = \frac{ \overline{\tilde{\mathbf{h}}^{H}_{k} \mathbf{w}^{\text{S}}_{k}}}{\sum\limits_{j=1}^{K}|\tilde{\mathbf{h}}^{H}_{k} \mathbf{w}^{\text{S}}_{j}|^{2} + |\mathbf{u}^{H}_{k} \mathbf{w}^{\text{B}}_{k,\text{c}}|^2 + \sum\limits_{j=1}^{L}|\mathbf{u}^{H}_{k} \mathbf{w}^{\text{B}}_{k,j}|^{2}+ P^{\text{E}}_{k}},
\end{equation}
\begin{equation}
	\label{update_mu2}
	\small
	\mu^{*}_{k,l} = \frac{\overline{\mathbf{v}^{H}_{k,l} \mathbf{w}^{\text{B}}_{k,l}}}{\sum\limits_{j=1}^{L}|\mathbf{v}^{H}_{k,l} \mathbf{w}^{\text{B}}_{k,j}|^{2} + |\tilde{\mathbf{f}}^{H}_{k,l} \mathbf{w}^{\text{S}}_{\text{c}}|^2 + \sum\limits_{j=1}^{K}|\tilde{\mathbf{f}}^{H}_{k,l} \mathbf{w}^{\text{S}}_{j}|^{2}+P^{\text{U}}_{k,l}},
\end{equation}
\begin{equation}
	\label{update_e_omega}
	e^{*}_{k,j} = e_{k,j}(\mathbf{\Upsilon},\mu^{*}_{k,j}),\omega^{*}_{k,j}=\frac{1}{e^{*}_{k,j}}.
\end{equation}

Then, we can optimize $\mathbf{\Upsilon}$ by solving the following problem.
\begin{align}
	(\text{P3}) \min_{\mathbf{\Upsilon}} &\sum\limits_{k=1}^{K} \sum\limits_{j=0}^{L} \alpha_{k,j} \omega^{*}_{k,j} e^{*}_{k,j} \tag{P3.a} \\
	\text{s.t.} &(\text{\ref{P2.b}}). \tag{P3.b}
\end{align}

\subsection{Optimizing $\{\mathbf{w}^{\textnormal{S}}, \mathbf{w}^{\textnormal{B}}\}$ Given $\{\mathbf{r}^{\textnormal{E}},\mathbf{r}^{\textnormal{U}},\mathbf{\Psi},\mathbf{R}^{\textnormal{R}}, \mathbf{t}^{\textnormal{R}}\}$}
The sub-problem (P3.1) can be formulated as follows.
\begin{align}
	(\text{P3.1}) \min_{\mathbf{w}^{\text{S}}, \mathbf{w}^{\text{B}}} &f(\mathbf{w}^{\text{S}}, \mathbf{w}^{\text{B}}) \tag{P3.1.a} \\
	\text{s.t.} &(\text{\ref{P1.c}})-(\text{\ref{P1.f}}),(\text{\ref{P1.k}}), \tag{P3.1.b}
\end{align}
where $f(\mathbf{w}^{\text{S}}, \mathbf{w}^{\text{B}})$ can be expanded as (\ref{P3.1.a}). In addition, the constraint (\ref{P1.k}) can be reformulated as
\begin{figure*}[t]
	\normalsize
	\begin{align}
		f(\mathbf{w}^{\text{S}}, \mathbf{w}^{\text{B}}) = &\sum\limits_{k=1}^{K} \alpha_{k,0} \omega_{k,0} (|\mu_{k,0}|^{2} (\sum_{j=1}^{K}|\tilde{\mathbf{h}}^{H}_{k} \mathbf{w}^{\text{S}}_{j}|^{2} + |\mathbf{u}^{H}_{k} \mathbf{w}^{\text{B}}_{k,\text{c}}|^2 + \sum_{j=1}^{L}|\mathbf{u}^{H}_{k} \mathbf{w}^{\text{B}}_{k,j}|^{2})-2 \text{Re}\{\mu_{k,0} \tilde{\mathbf{h}}^{H}_{k} \mathbf{w}^{\text{S}}_{k}\}) + \notag  \\
		&\sum\limits_{k=1}^{K} \sum\limits_{l=1}^{L} \alpha_{k,l} \omega_{k,l} (|\mu_{k,l}|^{2} (\sum_{j=1}^{L}|\mathbf{v}^{H}_{k,l} \mathbf{w}^{\text{B}}_{k,j}|^{2} + |\tilde{\mathbf{f}}^{H}_{k,l} \mathbf{w}^{\text{S}}_{\text{c}}|^2 + \sum_{j=1}^{K}|\tilde{\mathbf{f}}^{H}_{k,l} \mathbf{w}^{\text{S}}_{j}|^{2})-2 \text{Re}\{\mu_{k,l} \mathbf{v}^{H}_{k,l} \mathbf{w}^{\text{B}}_{k,l}\}) \label{P3.1.a}
	\end{align}
	\hrulefill
	\vspace*{-4pt}
\end{figure*}
\begin{equation}
	\label{P3.1.c}
	\hspace*{-0.6em}
	\resizebox{0.91\linewidth}{!}{$\Gamma^{\text{E}}_{k} (\sum\limits_{j \neq k}^{K}|\tilde{\mathbf{h}}^{H}_{k} \mathbf{w}^{\text{S}}_{j}|^{2} + |\mathbf{u}^{H}_{k} \mathbf{w}^{\text{B}}_{k,c}|^{2} + \sum\limits_{j=1}^{L}|\mathbf{u}^{H}_{k} \mathbf{w}^{\text{B}}_{k,j}|^{2} + P^{\text{E}}_{k}) \leq f^{\text{S}},$}
\end{equation}
\vspace*{-8pt}
\begin{equation}
	\label{P3.1.d}
	\hspace*{-0.6em}
	\resizebox{0.91\linewidth}{!}{$\Gamma^{\text{U}}_{k,l} (\sum\limits_{j \neq l}^{L}|\mathbf{v}^{H}_{k} \mathbf{w}^{\text{B}}_{k,j}|^{2} + |\tilde{\mathbf{f}}^{H}_{k,l} \mathbf{w}^{\text{S}}_{c}|^{2} + \sum\limits_{j=1}^{K}|\tilde{\mathbf{f}}^{H}_{k,l} \mathbf{w}^{\text{S}}_{j}|^{2} + P^{\text{U}}_{k,l}) \leq f^{\text{B}} ,$} 
\end{equation}
where $\Gamma^{\text{E}}_{k} = 2^{(R^{\text{E}}_{\text{min}} - r^{\text{E}}_{k})} - 1$ and $\Gamma^{\text{U}}_{k,l} = 2^{(R^{\text{U}}_{\text{min}} - r^{\text{U}}_{k,l})} - 1$. Moreover, the above $f^{\text{S}}$ and $f^{\text{B}}$ are the first-order Taylor expansions of 
$|\tilde{\mathbf{h}}^{H}_{k} \mathbf{w}^{\text{S}}_{k}|^2$ and 
$|\mathbf{v}^{H}_{k,l} \mathbf{w}^{\text{B}}_{k,l}|^2$ at point $\mathbf{w}^{\text{S}*}_{k}$ and $\mathbf{w}^{\text{B}*}_{k,l}$, respectively.
\begin{equation}
	\label{taylor_1}
	f^{\text{S}}=|\tilde{\mathbf{h}}^{H}_{k} \mathbf{w}^{\text{S}*}_{k}|^2+2\text{Re}\{\overline{\tilde{\mathbf{h}}^{H}_{k} \mathbf{w}^{\text{S}*}_{k}}\tilde{\mathbf{h}}^{H}_{k}(\mathbf{w}^{\text{S}}_{k}-\mathbf{w}^{\text{S}*}_{k}) \},
\end{equation}
\begin{equation}
	\label{taylor_2}
	f^{\text{B}}= |\mathbf{v}^{H}_{k,l} \mathbf{w}^{\text{B}*}_{k,l}|^2 + 2\text{Re}\{\overline{\mathbf{v}^{H}_{k,l} \mathbf{w}^{\text{B}*}_{k,l}}\mathbf{v}^{H}_{k,l}(\mathbf{w}^{\text{B}}_{k,l}-\mathbf{w}^{\text{B}*}_{k,l})\}.
\end{equation}

Additionally, defining $\Gamma^{\text{E,c}} \triangleq 2^{\sum_{k=1}^{K}r^{\text{E}}_{k}}-1$ and $\Gamma^{\text{U,c}}_{k} \triangleq 2^{\sum_{l=1}^{L}r^{\text{U}}_{k,l}}-1$, the constraints (\ref{P1.c})(\ref{P1.d}) can be written as
\begin{equation}
	\label{P3.1.e}
	\hspace*{-0.6em}
	\resizebox{0.91\linewidth}{!}{$\Gamma^{\text{E,c}} (\sum\limits_{j=1}^{K}|\tilde{\mathbf{h}}^{H}_{k} \mathbf{w}^{\text{S}}_{j}|^{2} + |\mathbf{u}^{H}_{k} \mathbf{w}^{\text{B}}_{k,c}|^{2} + \sum\limits_{j=1}^{L}|\mathbf{u}^{H}_{k} \mathbf{w}^{\text{B}}_{k,j}|^{2} + P^{\text{E}}_{k}) \leq f^{\text{S,c}},$}
\end{equation}
\vspace*{-12pt}
\begin{equation}
	\label{P3.1.f}
	\hspace*{-0.6em}
	\resizebox{0.91\linewidth}{!}{$\Gamma^{\text{U,c}}_{k} (\sum\limits_{j=1}^{L}|\mathbf{v}^{H}_{k,l} \mathbf{w}^{\text{B}}_{k,j}|^{2} + |\tilde{\mathbf{f}}^{H}_{k,l} \mathbf{w}^{\text{S}}_{c}|^{2} + \sum\limits_{j=1}^{K}|\tilde{\mathbf{f}}^{H}_{k,l} \mathbf{w}^{\text{S}}_{j}|^{2} + P^{\text{U}}_{k,l}) \leq f^{\text{B,c}},$}
	\vspace*{-0.6em}
\end{equation}
where $f^{\text{S,c}}$ and $f^{\text{B,c}}$ are the first-order Taylor expansions of $|\tilde{\mathbf{h}}^{H}_{k} \mathbf{w}^{\text{S}}_{\text{c}}|^2$ and $|\mathbf{v}^{H}_{k,l} \mathbf{w}^{\text{B}}_{k,\text{c}}|^2$ that are similar to above (\ref{taylor_1}) and (\ref{taylor_2}). At this point, the objective function (\ref{P3.1.a}) can be solved by the CVX toolbox.
\vspace*{-6pt}
\subsection{Optimizing $\mathbf{\Psi}$ Given $\{\mathbf{w}^{\textnormal{S}},\mathbf{w}^{\textnormal{B}},\mathbf{r}^{\textnormal{E}},\mathbf{r}^{\textnormal{U}},\mathbf{R}^{\textnormal{R}}, \mathbf{t}^{\textnormal{R}}\}$}
The sub-problem can be formulated as
\begin{align}
	(\text{P3.2}) \min_{\mathbf{\Psi}} &\text{ } g(\mathbf{\Psi}) \tag{P3.2.a} \\
	\text{s.t.} &(\text{\ref{P1.c}})(\text{\ref{P1.d}})(\text{\ref{P1.g}})(\text{\ref{P1.k}}), \tag{P3.2.b}
\end{align}
where $g(\mathbf{\Psi})$ is expanded as (\ref{P3.2.a}) and can be rearranged as
\begin{figure*}[t]
	\normalsize
	\vspace*{-12pt}
	\begin{align}
		g(\mathbf{\Psi})= &\sum\limits_{k=1}^{K} \alpha_{k,0} \omega_{k,0} (|\mu_{k,0}|^{2} \sum_{j=1}^{K}|(\mathbf{h}^{H}_k + \mathbf{g}^{H}_{k} \mathbf{\Psi}_{k} \mathbf{G}_k) \mathbf{w}^{\text{S}}_{j}|^{2} - 2 \text{Re}\{\mu_{k,0} (\mathbf{h}^{H}_k + \mathbf{g}^{H}_{k} \mathbf{\Psi}_{k} \mathbf{G}_k) \mathbf{w}^{\text{S}}_{k}\})+ \notag \\ &\sum\limits_{k=1}^{K} \sum\limits_{l=1}^{L} \alpha_{k,l} \omega_{k,l} |\mu_{k,l}|^{2} (|(\mathbf{f}^{H}_{k,l} + \mathbf{q}^{H}_{k,l} \mathbf{\Psi}_k \mathbf{G}_k) \mathbf{w}^{\text{S}}_{\text{c}}|^2 + \sum_{j=1}^{K}|(\mathbf{f}^{H}_{k,l} + \mathbf{q}^{H}_{k,l} \mathbf{\Psi}_k \mathbf{G}_k) \mathbf{w}^{\text{S}}_{j}|^{2}) \label{P3.2.a}
	\end{align}
	\hrulefill
	\vspace*{-16pt}
\end{figure*}
\begin{equation}
	\small
	\label{P3.2.a_simple}
	g(\boldsymbol{\zeta})=m + \sum\limits_{k=1}^{K} \sum\limits_{j=1}^{K+(K+1)L} |\boldsymbol{\zeta}_{k}^{T} \mathbf{c}_{k,j}|^2 +\sum\limits_{k=1}^{K} \sum\limits_{j=1}^{(K+1)(L+1)} \text{Re}\{\boldsymbol{\zeta}_{k}^{T} \mathbf{d}_{k,j}\},
\end{equation}
where $\boldsymbol{\zeta} \triangleq [\boldsymbol{\zeta}_{1},\boldsymbol{\zeta}_{2},...,\boldsymbol{\zeta}_{K}]$ and $m$ represents the complex constant unrelated to $\boldsymbol{\zeta}$. Furthermore, the constraints (\ref{P1.c})(\ref{P1.d})(\ref{P1.k}) can be reformulated as
\vspace*{-6pt}
\begin{align}
	\label{P3.3.c}
	\Gamma^{\text{E,c}} (\sum\limits_{j=1}^{K}|\mathbf{h}^{H}_k \mathbf{w}^{\text{S}}_{j} + \boldsymbol{\zeta}_{k}^{T}  \mathrm{diag}(\mathbf{g}^{H}_{k})\mathbf{G}_k \mathbf{w}^{\text{S}}_{j} |^{2} +|\mathbf{u}^{H}_{k} \mathbf{w}^{\text{B}}_{k,c}|^{2}\\
	+ \sum\limits_{j=1}^{L}|\mathbf{u}^{H}_{k} \mathbf{w}^{\text{B}}_{k,j}|^{2} + P^{\text{E}}_{k}) \leq g^{\text{S,c}}, \notag 
\end{align}
\vspace*{-16pt}
\begin{align}
	\label{P3.3.d}
	&\hspace*{-0.35em}\Gamma^{\text{U,c}}_{k} (\sum\limits_{j=1}^{L}|\mathbf{v}^{H}_{k,l} \mathbf{w}^{\text{B}}_{k,j}|^{2} + |\mathbf{f}^{H}_{k,l} \mathbf{w}^{\text{S}}_{\text{c}}+\boldsymbol{\zeta}_{k}^{T}  \mathrm{diag}(\mathbf{q}^{H}_{k,l})\mathbf{G}_k \mathbf{w}^{\text{S}}_{\text{c}}|^{2} \\ &\hspace*{-1.0em}+\sum\limits_{j=1}^{K}|\mathbf{f}^{H}_{k,l} \mathbf{w}^{\text{S}}_{j}+\boldsymbol{\zeta}_{k}^{T}  \mathrm{diag}(\mathbf{q}^{H}_{k,l})\mathbf{G}_k \mathbf{w}^{\text{S}}_{j}|^{2} + P^{\text{U}}_{k,l}) \leq |\mathbf{v}^{H}_{k,l} \mathbf{w}^{\text{B}}_{k,\text{c}}|^2, \notag
\end{align}
\vspace*{-16pt}
\begin{align}
	\label{P3.3.e}
	\Gamma^{\text{E}}_{k} (\sum\limits_{j \neq k}^{K}|\mathbf{h}^{H}_k \mathbf{w}^{\text{S}}_{j} + \boldsymbol{\zeta}_{k}^{T}  \mathrm{diag}(\mathbf{g}^{H}_{k})\mathbf{G}_k \mathbf{w}^{\text{S}}_{j}|^{2} + |\mathbf{u}^{H}_{k} \mathbf{w}^{\text{B}}_{k,c}|^{2} \\ +\sum\limits_{j=1}^{L}|\mathbf{u}^{H}_{k} \mathbf{w}^{\text{B}}_{k,j}|^{2} + P^{\text{E}}_{k}) \leq g^{\text{S}}, \notag 
\end{align}
\vspace*{-16pt}
\begin{align}
	\label{P3.3.f}
	&\hspace*{-0.35em}\Gamma^{\text{U}}_{k,l} (\sum\limits_{j \neq l}^{L}|\mathbf{v}^{H}_{k,l} \mathbf{w}^{\text{B}}_{k,j}|^{2} + |\mathbf{f}^{H}_{k,l} \mathbf{w}^{\text{S}}_{\text{c}}+\boldsymbol{\zeta}_{k}^{T}  \mathrm{diag}(\mathbf{q}^{H}_{k,l})\mathbf{G}_k \mathbf{w}^{\text{S}}_{\text{c}}|^{2} \\
	&\hspace*{-1.0em}+\sum\limits_{j=1}^{K}|\mathbf{f}^{H}_{k,l}\mathbf{w}^{\text{S}}_{j}+\boldsymbol{\zeta}_{k}^{T}  \mathrm{diag}(\mathbf{q}^{H}_{k,l})\mathbf{G}_k \mathbf{w}^{\text{S}}_{j}|^{2} + P^{\text{U}}_{k,l}) \leq |\mathbf{v}^{H}_{k,l} \mathbf{w}^{\text{B}}_{k,l}|^2 ,\notag
\end{align}
where $g^{\text{S,c}}$ and $g^{\text{S}}$ are the first-order Taylor expansions of $|\tilde{\mathbf{h}}^{H}_{k} \mathbf{w}^{\text{S}}_{\text{c}}|^2$ and $|\tilde{\mathbf{h}}^{H}_{k} \mathbf{w}^{\text{S}}_{k}|^2$ at $\boldsymbol{\zeta}^{*}$. Additionally, (\ref{P1.g}) can be rewritten as
\begin{equation}
	\label{module_1}
	|[\boldsymbol{\zeta}_{k}]_{i}|=1,\forall k \in \mathcal{K},i=1,2,...,N_{\text{R}}.
\end{equation}

Using alternating direction method of multipliers (ADMM) to handle (\ref{module_1}) with a positive penalty parameter $\rho$ \cite{ADMM}, (P3.2) can be reformulated as 
\begin{align}
	(\text{P3.3}) \min_{\boldsymbol{\zeta},\mathbf{Z},\mathbf{Y}} 
	&g(\boldsymbol{\zeta})+\frac{\rho}{2}\|\boldsymbol{\zeta}-\mathbf{Z}+\mathbf{Y}\|^2-\frac{\rho}{2}\|\mathbf{Y}\|^2 \label{P3.3.a} \tag{P3.3.a} \\
	\text{s.t.} &(\text{\ref{P3.3.c}})(\text{\ref{P3.3.d}})(\text{\ref{P3.3.e}})(\text{\ref{P3.3.f}}), \label{P3.3.b} \tag{P3.3.b} \\
	&\boldsymbol{\zeta} = \mathbf{Z}, \label{P3.3.c2} \tag{P3.3.c} \\
	&[|\mathbf{Z}|]_{k,i} = 1,\forall k \in \mathcal{K},i=1,2,...,N_{\text{R}}. \label{P3.3.d2} \tag{P3.3.d}
\end{align}

In each iteration of ADMM, the update methods for each variable are as follows:
\vspace*{-6pt}
\begin{equation}
	\label{updatezeta}
	\boldsymbol{\zeta}^{*} = \arg \min_{\boldsymbol{\zeta}} g(\boldsymbol{\zeta})+\frac{\rho}{2}\|\boldsymbol{\zeta}-\mathbf{Z}+\mathbf{Y}\|^2,
	\vspace*{-6pt}
\end{equation}
\begin{equation}
	\label{updateZ}
	\mathbf{Z}^{*} = \arg \min_{[|\mathbf{Z}|]_{k,i} = 1} \|\boldsymbol{\zeta}^{*}-\mathbf{Z}+\mathbf{Y}\|^2,
	\vspace*{-6pt}
\end{equation}
\begin{equation}
	\mathbf{Y}^{*} = \mathbf{Y} + \boldsymbol{\zeta}^{*} - \mathbf{Z}^{*}.
	\vspace*{-6pt}
\end{equation}
\vspace*{-4pt}
When updating $\mathbf{Z}$, there is an analytical solution as
\begin{equation}
	[\mathbf{Z}]_{k,i} = \frac{[\boldsymbol{\zeta}^{*}]_{k,i}+[\mathbf{Y}]_{k,i}}{|[\boldsymbol{\zeta}^{*}]_{k,i}+[\mathbf{Y}]_{k,i}|}.
\end{equation}

The iteration terminates when both the primal and dual residuals, expressed as $\| \boldsymbol{\zeta} - \mathbf{Z}  \|$ and $ \rho \| \mathbf{Z} - \mathbf{Z}^{*} \|$ , satisfy the required accuracy.

\subsection{Optimizing $\mathbf{R}^{\textnormal{R}}, \mathbf{t}^{\textnormal{R}}$ Given $\{\mathbf{w}^{\textnormal{S}},\mathbf{w}^{\textnormal{B}},\mathbf{\Psi},\mathbf{r}^{\textnormal{E}},\mathbf{r}^{\textnormal{U}}\}$}
\begin{figure*}[!t]
	\centering
	\begin{subfigure}[b]{0.32\textwidth}
		\centering
		\includegraphics[width=2.4in]{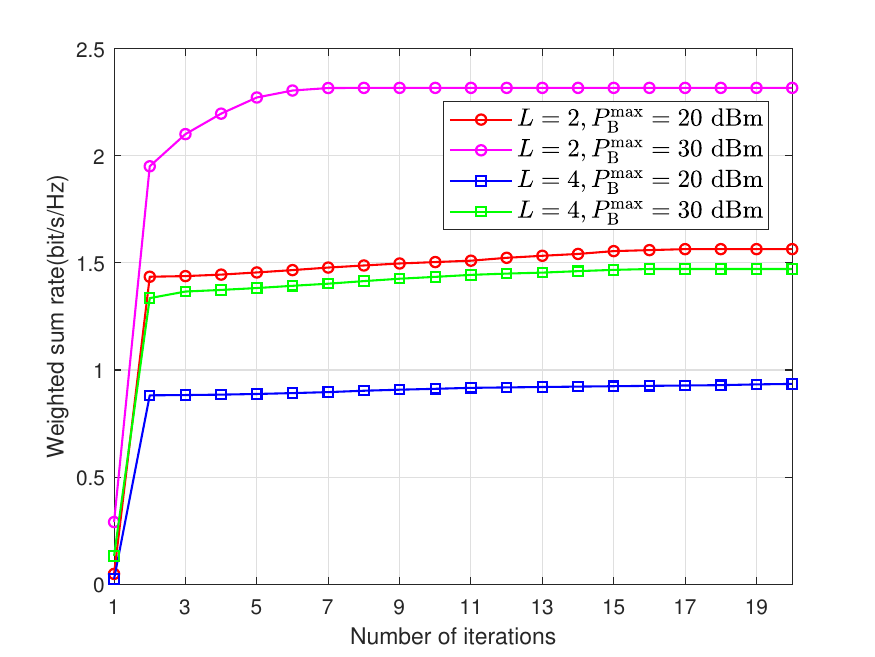}
		\caption{Weighted sum rate versus iteration numbers.}
		\label{simu1}
	\end{subfigure}
	\hfill
	\begin{subfigure}[b]{0.32\textwidth}
		\centering
		\includegraphics[width=2.4in]{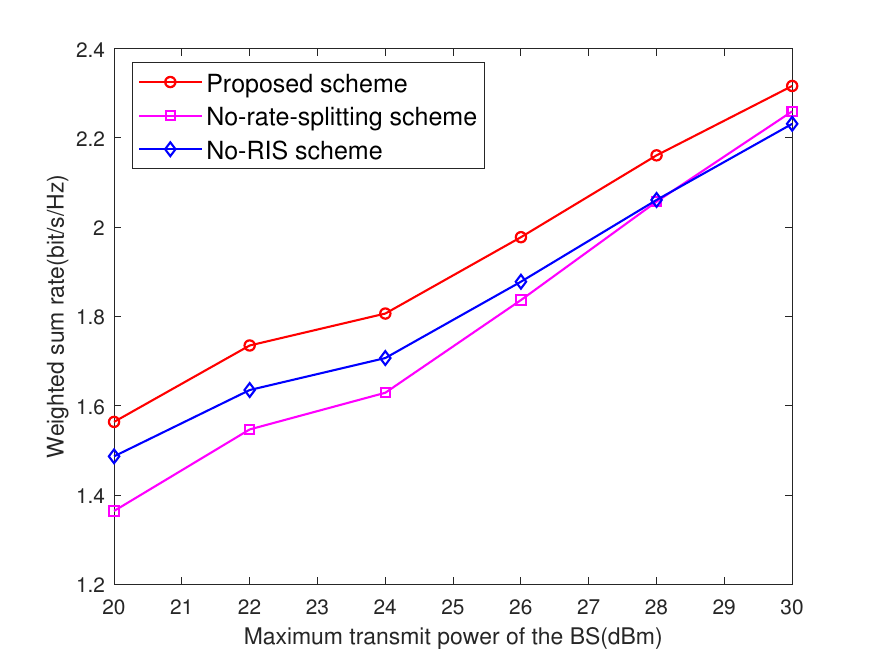}
		\caption{Weighted sum rate versus maximum transmit power of the BS.}
		\label{simu2}
	\end{subfigure}
	\hfill
	\begin{subfigure}[b]{0.32\textwidth}
		\centering
		\includegraphics[width=2.4in]{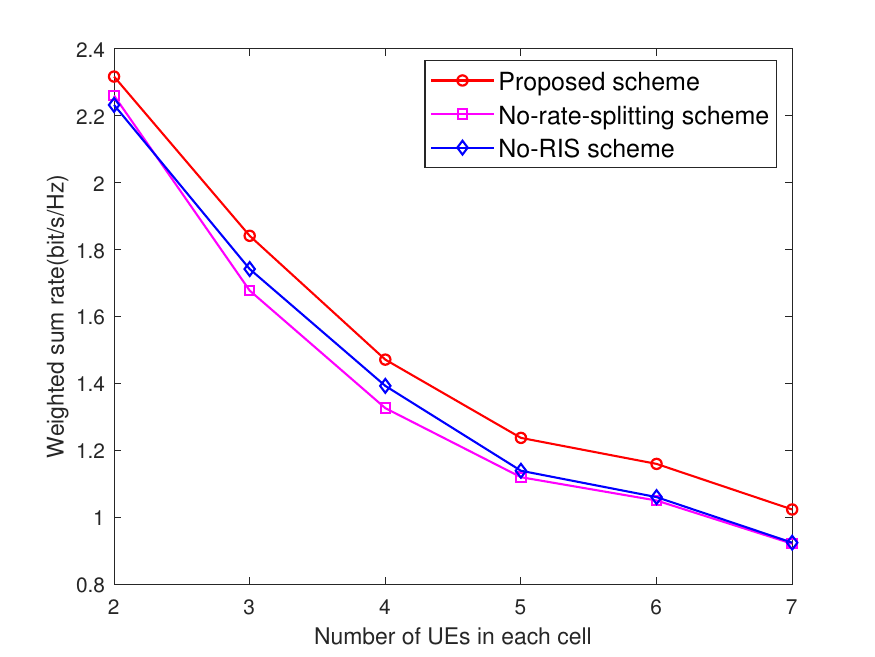}
		\caption{Weighted sum rate versus number of UEs in each cell.}
		\label{simu3}
	\end{subfigure}
	\caption{Performance evaluation of the proposed algorithm}
	\vspace*{-10pt}
	\label{fig:three_scenarios}
\end{figure*}
Given any feasible $\{\mathbf{w}^{\textnormal{S}},\mathbf{w}^{\textnormal{B}},\mathbf{r}^{\textnormal{E}},\mathbf{r}^{\textnormal{U}},\mathbf{\Psi}\}$ and omitting variables unrelated to $\mathbf{R}^{\textnormal{R}}$ and $\mathbf{t}^{\textnormal{R}}$, the sub-problem can be formulated as
\vspace*{-6pt}
\begin{align}
	(\text{P3.4}) \min_{\mathbf{R}^{\textnormal{R}},\mathbf{t}^{\textnormal{R}}} &h(\mathbf{R}^{\textnormal{R}},\mathbf{t}^{\textnormal{R}}) \tag{P3.4.a} \\
	\text{s.t.} &(\text{\ref{P1.c}})(\text{\ref{P1.d}}),(\text{\ref{P1.h}})-(\text{\ref{P1.k}}), \tag{P3.4.b}
\end{align}
where the objective function $h(\mathbf{R}^{\textnormal{R}},\mathbf{t}^{\textnormal{R}})$ takes the same form as $g(\mathbf{\Psi})$. It is difficult to design a low-complexity algorithm to optimize $\mathbf{R}^{\textnormal{R}}$ and $\mathbf{t}^{\textnormal{R}}$. However, when the UAV's attitude and position accuracy are limited or not strictly required, it is feasible to design an exhaustive search algorithm.

First, we reduce the dimensionality of the optimization variables. According to (\ref{P1.h}), we can process $\mathbf{R}^{\textnormal{R}}_{k}$ using the Euler angle decomposition as
\vspace*{-4pt}
\begin{equation}
	\mathbf{R}^{\textnormal{R}}_{k} = \mathbf{R}^{\textnormal{R}}_{k,x}(\beta_{k,x})\mathbf{R}^{\textnormal{R}}_{k,y}(\beta_{k,y})\mathbf{R}^{\textnormal{R}}_{k,z}(\beta_{k,z}),
	\vspace*{-4pt}
\end{equation}
where $\beta_{k,x}$, $\beta_{k,y}$ and $\beta_{k,z}$ represent the rotation angles around the $x$, $y$ and $z$-axes, respectively. Thus, we only need to traverse the three rotation angles to optimizing $\mathbf{R}^{\textnormal{R}}_{k}$.

Then, according to (\ref{P1.i}), we represent the central axis of the $k$-th parabolic antenna as a straight line in $\mathcal{S}_0$. Therefore, the coordinates of the RIS can be expressed as
\begin{equation}
	\mathbf{Q}^{\text{R},\mathcal{S}_{0}}_{k} = \mathbf{t}^{\text{R}}_{k} = \mathbf{Q}^{\text{E},\mathcal{S}_{0}}_{k} + b_{k} \mathbf{p}_{k}, \forall k \in \mathcal{K},
\end{equation}
where $\mathbf{p}_{k}$ denotes the unit direction vector of the $k$-th central axis, and $b_{k}$ determines the position of the RIS along the central axis. Furthermore, $b_{k}$ should satisfy the constraint as
\begin{equation}
	- H_{\text{min}}/[\mathbf{p}_{k}]_{3} \leq b_{k} \leq - H_{\text{max}}/[\mathbf{p}_{k}]_{3}, \forall k \in \mathcal{K}.
\end{equation}

By following this alternating optimization procedure, the optimal solution to Problem (P3.4) can be obtained.
\subsection{Complexity Analysis}
It is observed that the complexity of the greedy algorithm for solving (P1.1) is linear. In addition, (P3.1) is a quadratically constrained quadratic program (QCQP), which can be transformed into a second-order cone programming (SOCP). The computational complexity of solving an SOCP depends on the number of the second-order cone (SOC) constraints $k_\text{soc}$, the dimension of the optimization variables $m_\text{soc}$, and the dimension of the SOC constraints $n_\text{soc}$, and can be expressed as $O(k_\text{soc}^{1/2}(m_\text{soc}^3+m_\text{soc}^{2}\sum_{i=1}^{k_\text{soc}}n_\text{soc}+\sum_{i=1}^{k_\text{soc}}n_\text{soc}^{2}))$ \cite{complexity}.
Therefore, The computational complexity of solving (P3.1) can be approximately expressed as $O(K^{7/2}L^{1/2}(N_{\text{S}}+L N_{\text{B}})^3\log(1/\epsilon_{\text{C}}))$, with $\epsilon_{\text{C}}$ denoting the target optimization precision, when $N_{\text{S}}, N_{\text{B}} \gg K, L$. Similarly, since the updates of $\mathbf{Z}$ and $\mathbf{Y}$ have closed-form solutions, the computational complexity of solving (P3.3) mainly depends on the update of $\boldsymbol{\zeta}$, and can be approximately expressed as $O(K^{7/2}L^{1/2}N_{\text{R}}^{3}\log(1/\epsilon_{\text{D}}))$, when $N_{\text{R}} \gg K,L$. Moreover, when solving (P3.4), the precision of distance and angle affect the complexity, which is given by $O((1 / \epsilon_{\text{E}1})^3 \cdot 1 / \epsilon_{\text{E}2})$.
\section{NUMERICAL RESULTS}
\subsection{Parameter Setup and Benchmark Schemes}
Unless specified otherwise, the parameters of the simulations are set as follows. The satellite provides services to the ESs in $K=3$ cells. Meanwhile, in each cell, the BS serves $L=2$ UEs for data transmission.
Both the BS and the satellite are equipped with $N_{\text{S}}=N_{\text{B}}=8 \times 8 = 64$ antennas and operate at a frequency of $f=28$ GHz.
In addition, the RIS-UAV is equipped with $N_{\text{R}}=8 \times 8 = 64$ reflecting elements and  the inter-antenna spacing of all UPAs is set to half of the wavelength, i.e., $d_0=\lambda/2$. The maximum transmit powers of the BS and the satellite are set to $P^{\text{B}}_{\text{max}}=30$ dBm and $P^{\text{S}}_{\text{max}}=5$ W, respectively. For simplicity, the noise powers at both the ESs and the UEs are set to $P^{\text{E}}_{k},P^{\text{U}}_{k,l}=-80 \text{ dBm}, \forall k \in \mathcal{K}, l\in \mathcal{L}.$ 
Moreover, the parameters related to rain attenuation are set to $\mu = -3.125$ and $\sigma = 1.591$.

Under various conditions, we compared the proposed algorithm with two benchmark schemes. \textit{1) No-rate-splitting scheme}: This benchmark scheme is based on the proposed algorithm but excludes the rate-splitting step, aiming to evaluate the system performance in the absence of the common stream. \textit{2) No-RIS scheme}: This benchmark scheme is designed to evaluate the system performance without RIS assistance.
\subsection{Performance Analysis}
Fig.\ref{simu1} illustrates the convergence behaviour of the proposed algorithm under various conditions. It can be observed that the algorithm converges within a finite number of iterations under all settings. Moreover, An increase in the number of UEs or a reduction in transmit power lead to a decrease in the weighted sum rate, primarily due to the reduction in the SINR. Specifically, an increase in the number of UEs not only intensifies interference but also implies that the same transmit power must be distributed among more UEs.

Fig.\ref{simu2} and Fig.\ref{simu3} illustrate the trends of the weighted sum rate for the three schemes as the BS transmit power and the number of UEs vary, respectively. It can be observed that the weighted sum rate of all three schemes increases with higher transmit power or fewer UEs. In addition, the proposed scheme outperforms the other two schemes, demonstrating that rate splitting and RIS-UAV can provide significant performance gains in this system.
\section{CONCLUSIONS}
In this paper, we proposed an RIS-UAV-assisted approach to enhance the performance of SAGIN with RSMA, and formulated a joint optimization problem involving beamforming, RIS phase shifts, and the UAV’s position and orientation. Moreover, we developed a two-layer iterative algorithm, where the original problem was decomposed into multiple sub-problems in the outer loop. Furthermore, we designed solution methods for each sub-problem. Numerical results validate the convergence of the proposed algorithm and demonstrate its performance advantages compared with other schemes.

\bibliographystyle{IEEEtran}
\bibliography{references}

\vspace{12pt}
\end{document}